# Two-Dimensional Si-Ge Monolayers: Stabilities, Structures and Electronic Properties

Jiating Lu[1,2], Xi Zhang[1,2,*], Limeng Shen[1,2], Ya Nie [1,2] and Gang Xiang [1,2,*]

[1] College of Physics, Sichuan University, Chengdu, 610064, China

[2] Key Laboratory of Radiation Physics and Technology and Key Laboratory of High Energy Density Physics and Technology of Ministry of Education, Sichuan University, Chengdu, 610064, China

**Abstract**

Si-Ge monolayers (SiGeM) with different elementary proportion x (0<x<1) were systematically studied for the first-time using ab initio calculations in this work. The structural stabilities of the $Si_{1-x}Ge_xM$ with different symmetries were investigated using phonon spectra, and an infinite miscibility between Si and Ge elements were revealed in the 2D honeycomb structures. The simulated scanning tunneling microscope images and Raman and infrared active modes of the $Si_{1-x}Ge_xM$ were then obtained for structural characterizations. Interestingly, the study of electronic properties revealed not previously reported oscillatory nonlinear dependence of band gap values on the elementary proportion x in the $Si_{1-x}Ge_xM$, which suggests an alternative way for tuning the band gaps of 2D materials. Additionally, low effective masses (0.008m0 ~ 0.021m0) of the carriers in the semiconducting $Si_{1-x}Ge_xM$ were found, which has potentials for high-speed applications. Considering the advantage of their compatibility with current Si-based technology and the trend of miniature of electronic devices, the $Si_{1-x}Ge_xM$ with stable structures and excellent properties would be important for 2D applications based on group IV materials.

*Corresponding authors: Xi Zhang, xizhang@scu.edu.cn；Gang Xiang, gxiang@scu.edu.cn

# INTRODUCTION

The saga of semiconductor has prevailed since 1947. Nowadays, most of the discrete electronic devices and integrated circuits are made of silicon (Si) bulk because of its abundance in the earth crust and superior physical properties including low band gap, high mobility and robust tunability by doping. Meanwhile, germanium (Ge), another important group IV element which has infinite miscibility with Si in three-dimensional (3D) SiGe alloys, provide important versatilities and more functionalities. In fact, $Si_{1-x}Ge_x$ (0<x<1) alloys have been widely used for high-speed semiconductor transistors.[1,2] Probably stimulated by the great success of Si-based semiconductors in 3D, the two-dimensional (2D) counterparts of group IV elements have been explored for decades. As early as in 1994, K. Takeda and K. Shiraishi first proposed that Si and Ge could form corrugated honeycomb structures.[3] This work did not attract much attention until graphene was discovered[4] and 2D materials generated a lot of interest. Nowadays, 2D honeycomb structures of Si atoms and Ge atoms, i.e., silicene and germanene, respectively, have been extensively studied.[5-35] Theoretical investigations have revealed interesting properties of silicene and germanene such as band topology,[6,7] quantum Hall effects,[6] quantum anomalous Hall effect,[8] valley-spin polarization,[9] band gap tuning by biaxial strain and electric field,[10] and chemical functionalization.[11] Experimental synthesis of silicene and germanene has also been reported. For instance, silicene have been grown on Ag (110) surface,[12] $ZrB_2$ (0001) thin films on Si substrates,[13] Ag (111) surface,[14-21] Ru surface,[22] Ir (111) substrate[23] and graphite.[24] Germanene have been fabricated on Pt (111) surface,[25,26] Au (111) surface,[27] Ag (111) surface,[28,29] Sb(111) surface[30], $MoS_2$ Surface,[31,32] Ge(110) surface[33] and Cu (111) surface.[34] Atomic-level silicene field-effect transistors have also been demonstrated.[35,36] However, most of the work has been focused on silicene and/or germanene, but not the binary honeycomb structures of Si and Ge atoms, i.e., Si-Ge monolayers (SiGeM). The $Si_{1-x}Ge_xM$ with different elementary proportion x (0<x<1) have been relatively unexplored. The stabilities and miscibility, the structural characteristics and the electronic properties of the 2D $Si_{1-x}Ge_xM$ still remain unknown.

In this work, $Si_{1-x}Ge_xM$ (0<x<1) were systematically explored for the first-time using *ab initio* calculations. The structural stabilities of the $Si_{1-x}Ge_xM$ with different symmetries were studied and the first 2D binary system with an infinite miscibility between two elements (Si and Ge) was revealed in the honeycomb structures. Then, the simulated scanning tunneling microscope (STM) images and Raman and infrared active modes of the $Si_{1-x}Ge_xM$ were calculated for future experimental verifications. Finally, the electronic properties of the $Si_{1-x}Ge_xM$ were investigated.

## COMPUTATIONAL METHODS

Our density-functional theory (DFT) based first principles calculations were performed by Vienna *ab initio* simulation package (VASP).[37,38] The projector augmented wave (PAW) potentials were adopted to describe the core electrons. The generalized gradient approximation (GGA) of Perdew, Burke, and Ernzernhof (PBE)[39] and PBE with the spin-orbit coupling (PBE+SOC) were adopted to describe the exchange and correlation potentials, respectively. The cutoff energy for expansion of the wavefunction into plane waves is set to be 500 eV in all simulations. In the calculations of the self-consistent field potential and total energy, we use a set of (25×25×1) k-point sampling to carry out Brillouinzone (BZ) integral in K space. Here the k-point mesh is generated by Monkhorst-Pack scheme.[40] The vacuum layer of 15 Å is selected to avoid the interlayer interaction of neighboring supercells. The tolerance for electron convergence was set as 1.0E-6eV, and the force was converged within -1.0E-3eV/ Å, so the atomic position and cell shape of experimental data were fully relaxed. We used VASP and density functional perturbation theory (DFPT) implemented in PHONONPY package[41] to get the phonon dispersion, vibration modes and the Raman frequency at the Γ point. To obtain the force constants for the phonon spectra calculations, atomic displacements of 0.01 Å were employed.

## RESULTS AND DISCUSSIONS

In order to explore the structural stabilities of the $Si_{1-x}Ge_xM$, the symmetries of the low-buckled structures composed of Si and Ge atoms were first studied. Starting from a 2×2×1 supercell of silicene with 8 atoms (see Figure 1), when a Si atom in the supercell is replaced by a Ge atom, there are 3 possible configurations and the Ge concentration is 12.5%. Keeping on doing this step by step, we can obtain totally 86 possible configurations and 7 different Ge concentrations. The SiGeM with different Ge concentrations are $Si_{0.875}Ge_{0.125}M$, $Si_{0.75}Ge_{0.25}M$, $Si_{0.625}Ge_{0.375}M$, $Si_{0.5}Ge_{0.5}M$, $Si_{0.375}Ge_{0.625}M$, $Si_{0.25}Ge_{0.75}M$ and $Si_{0.125}Ge_{0.875}M$, respectively. Symmetry analysis shows that $Si_{0.875}Ge_{0.125}M$ and $Si_{0.125}Ge_{0.875}M$ have *P3m1* symmetry, $Si_{0.75}Ge_{0.25}M$ and $Si_{0.25}Ge_{0.75}M$ have *C2/m*, *Pm* and *P-3m1*($D_{3d}$ point group:$C_3$, $3C_2$, $3\sigma$, $i$) symmetries, $Si_{0.625}Ge_{0.375}M$ and $Si_{0.375}Ge_{0.625}M$ have *Cm* and *P3m1*($C_{3v}$ point group:$C_3$, $3\sigma$) symmetries, and $Si_{0.5}Ge_{0.5}M$ have the most abundant symmetries of *P21/m*, *Cm*, *C2*, *P2/m* and *P3m1*. Among all the symmetries, *P3m1* and *P-3m1* have the highest symmetry and correspond to the case of nearly uniformly doped structures with highest entropy.

All the structures were optimized and the cohesive energies ($E_c$) of the $Si_{1-x}Ge_xM$ were calculated using the expression $E_c = E_{T[SiGe]} - E_{Si} - E_{Ge}$, where $E_{T[SiGe]}$ is the total energy of the optimized structure, and $E_{Si}$ and $E_{Ge}$ are the energies of free Si and Ge atoms, respectively. All the energies of the $Si_{1-x}Ge_xM$ with different symmetries were obtained and shown in Figure 1. All the energies are negative, indicating that the $Si_{1-x}Ge_xM$ are energetically favored. Additionally, for the same Ge doping concentration the energy variations of the $Si_{1-x}Ge_xM$ between different symmetries are small.

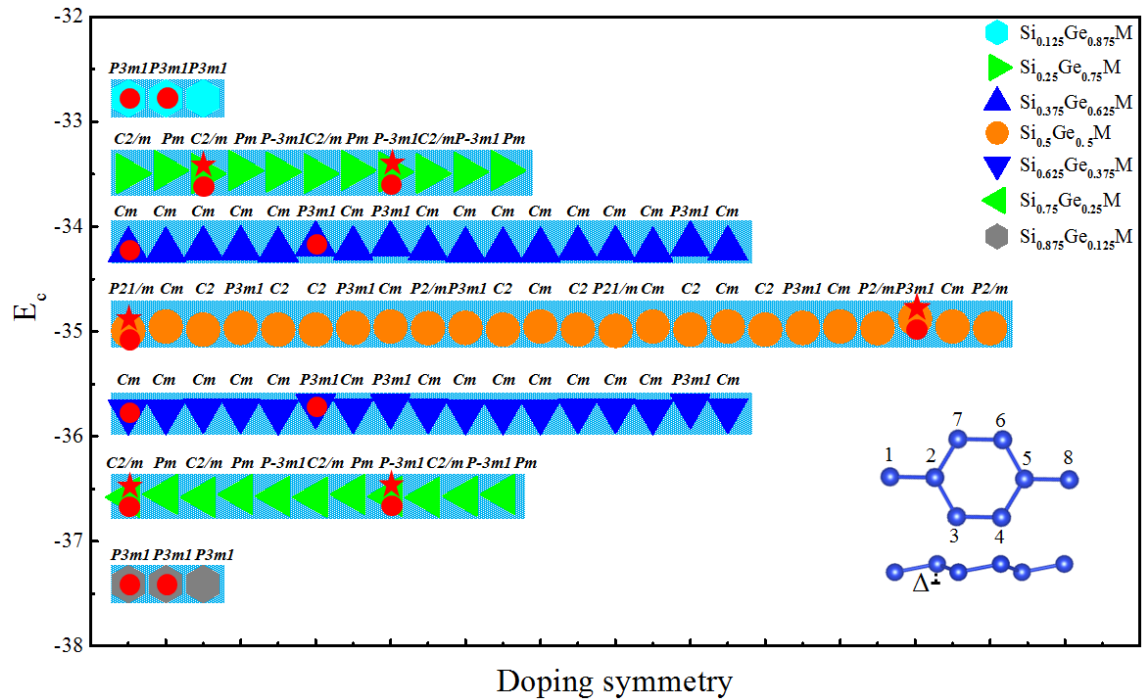

**FIG. 1**. Cohesive energy $E_c$ of the $Si_{1-x}Ge_xM$ as the function of doping symmetries. The inset in the right down corner shows the 2×2×1 supercell of silicene.

The structural stabilities of the $Si_{1-x}Ge_xM$ with different symmetries were then explored using phonon dispersions. For comparison, we also calculated the phonon spectra of silicene and germanene (not shown), and the results were consistent with previous theoretical work.[5,42-44]

Typical phonon spectra of $Si_{0.25}Ge_{0.75}M$, $Si_{0.5}Ge_{0.5}M$ and $Si_{0.25}Ge_{0.75}M$ along the high symmetric points in the Brillouin zone are shown in Figure 2a-f. For all the 2D honeycomb structures, there are three acoustical branches: LA, TA and ZA. Because of the fast attenuation of transverse force, when k→0, the LA and TA branches which contribute a lot to heat conduction are linear, while the dispersion of ZA branches which contribute little to thermal conductivity is quadratic. Most importantly, one can see that all the vibration frequencies calculated are positive without imaginary part, especially the frequencies of acoustical modes (ZA, TA and LA) as k→0, which ensures structural stability.[45] The results indicate that the $Si_{1-x}Ge_xM$ are dynamically stable at ground state even for long-wavelength lattice vibrations and are likely to be synthesized experimentally. The interaction between Si and Ge atoms can be further analyzed by electron localization function (ELF),[46] as shown in Figure 2g. Here, the ELF values vary from 0.53 for free electrons to 1 for fully localized electrons, and the values between 0.7 and 0.8 indicate covalent bonding character. Obviously, there are charge accumulations between Si-Si, Si-Ge and Ge-Ge bonds, indicating strong covalent interactions.

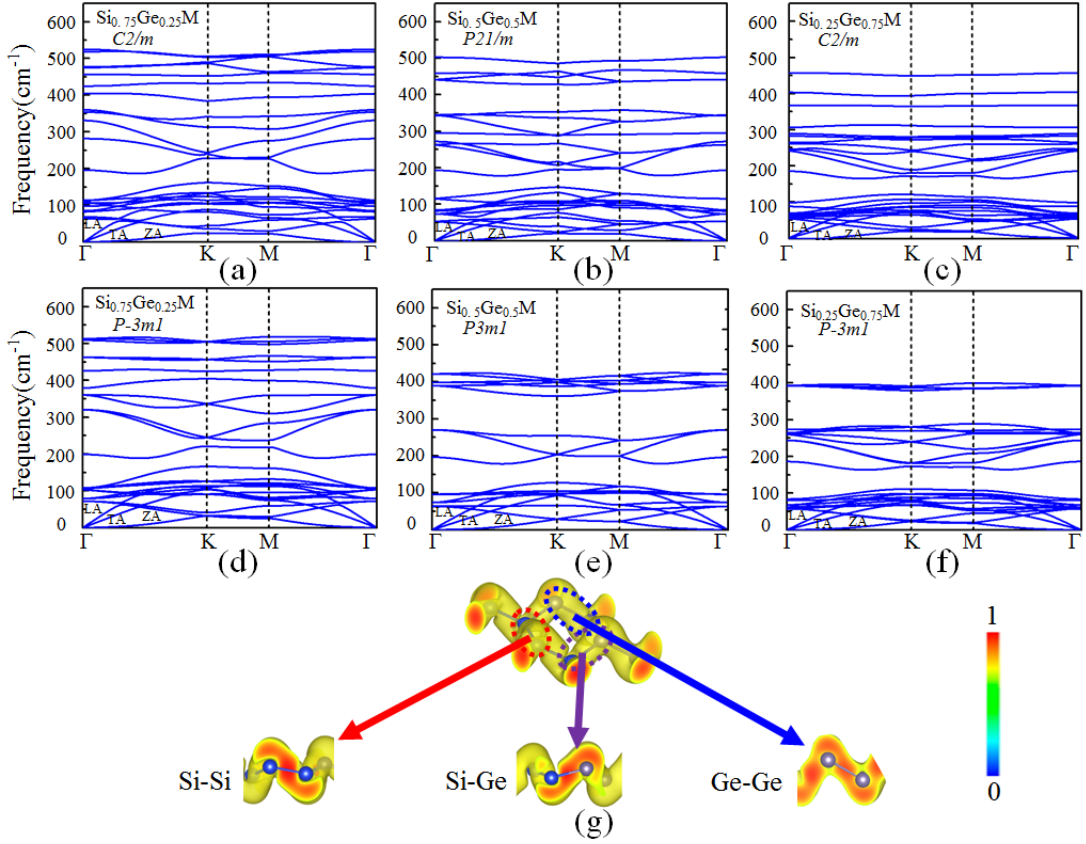

**FIG. 2**. (a- f) Typical phonon dispersion curves. (a) and (d) $Si_{0.75}Ge_{0.25}M$ with C2/m and *P-3m1* symmetry, (b) and (e) $Si_{0.5}Ge_{0.5}M$ with *P21/m* and *P3m1* symmetry, (c) and (f) $Si_{0.25}Ge_{0.75}M$ with *C2/m* and *P-3m1* symmetry, respectively. All of them are marked with red pentagrams in Figure 1. All the vibration frequencies calculated are positive without imaginary part. (g) the electronic local function (ELF) density diagrams of $Si_{0.5}Ge_{0.5}M$ with P21/m symmetry and its Si-Si bond, Si-Ge bond and Ge-Ge bond are given.

In short, our results reveal that the $Si_{1-x}Ge_xM$ (0<x<1) with all the elementary proportions x studied are structurally stable, which suggests an infinite miscibility between the Si and Ge atoms in the 2D honeycomb monolayers, just as those in 3D face-centered diamond-cubic SiGe crystals. To the best of our knowledge, this is the first 2D binary system which exhibits infinite miscibility of two elements and could provide a lot of varieties and functionalities in structures and properties for 2D applications.

**TABLE 1.** The structural parameters of the $Si_{1-x}Ge_xM$ with LE and HS doping strategies, where *a* represents the 2D hexagonal lattice constant, *d* the nearest-neighbor distance of Si-Si and Ge-Ge atoms (the bond length of Si-Ge bond is indicated in parentheses), *θ* the bond angle of adjacent bonds and $\triangle$ the buckling parameter.

| Name | *a*(Å) | | *d*(Å) | | *θ*(deg) | | $\triangle$(Å) | |
|---|---|---|---|---|---|---|---|---|
| | LE | HS | LE | HS | LE | HS | LE | HS |
| Silicene | 7.74 | 7.74 | 2.28 | 2.28 | 116.2 | 116.2 | 0.449 | 0.449 |
| $Si_{0.875}Ge_{0.125}M$ | 7.77 | 7.77 | 2.28(2.34) | 2.28(2.34) | 114.8 | 116.2 | 0.554 | 0.564 |
| $Si_{0.75}Ge_{0.25}M$ | 7.81 | 7.82 | 2.29(2.34) | 2.28(2.35) | 115.1 | 115.2 | 0.581 | 0.596 |
| $Si_{0.625}Ge_{0.375}M$ | 7.86 | 7.86 | 2.31(2.35) | 2.30(2.35) | 115.5 | 115.4 | 0.69 | 0.602 |
| $Si_{0.5}Ge_{0.5}M$ | 7.91 | 7.9 | 2.31(2.36) | (2.36) | 117.5 | 114 | 0.697 | 0.59 |
| $Si_{0.375}Ge_{0.625}M$ | 7.96 | 7.96 | 2.4(2.38) | 2.42(2.36) | 112.2 | 112.1 | 0.721 | 0.691 |
| $Si_{0.25}Ge_{0.75}M$ | 7.82 | 8 | 2.29(2.34) | 2.43(2.36) | 115.2 | 112.8 | 0.668 | 0.664 |
| $Si_{0.125}Ge_{0.875}M$ | 8.06 | 8.06 | 2.44(2.38) | 2.43(2.38) | 113.3 | 111.8 | 0.713 | 0.716 |
| Germanene | 8.12 | 8.12 | 2.44 | 2.44 | 112.4 | 112.4 | 0.689 | 0.689 |

In order to explore the structural characteristics of the $Si_{1-x}Ge_xM$, the lattice parameters were then calculated and shown in Table 1. Here, two doping strategies were considered: lowest energy (LE) strategy and highest entropy (HS) strategy. The LE doping strategy corresponds to the most energetically probable situation in equilibrium where the minimum energy principle rules, while the HS strategy corresponds to the situation where dopants are most evenly distributed and the doped system has the highest entropy, which also occurs experimentally in the synthesis of alloys. In both situations, as the Ge concentrations increases, the 2D hexagonal lattice constant (*a*) increases, the nearest-neighbor distance (*d*) of the Si-Si and Ge-Ge atoms increases, the bond angle ($\theta$) of adjacent bonds decreases, and the buckling parameter ($\triangle$) increases, while the bond length of Si-Ge bond change a little (from 2.34 to 2.38Å). The larger bond length (2.28-2.44Å) of the Si-Si and Ge-Ge atoms in the SiGeM relative to that of C-C atoms in graphene (~1.42Å) prevents the formation of strong π bonds between atoms, leading to the formation of buckling structure of the SiGeM. Consequently, the bucking structure further results in mixed $sp^2$-$sp^3$ hybridization. According to the Jahn-Teller theorem,[47] the total energy of the structure is reduced and the structural stability is regained.[48]

The simulated STM images of the $Si_{1-x}Ge_xM$ with different doping strategies under 0.5V bias were shown in Figure 3. The bright and dark dots represent the top and bottoms, respectively. The area surrounded by the blue rhombus represents the location of the primitive cell. Obvious honeycomb structures can be observed from STM imaging. For the LE doping strategy, the brightness of the atoms in $Si_{0.875}Ge_{0.125}M$ and $Si_{0.125}Ge_{0.875}M$ is basically uniform, while the brightness of the atoms in the others is different, indicating that $Si_{0.875}Ge_{0.125}M$ and $Si_{0.125}Ge_{0.875}M$ are flatter than the other SiGeM. This was more intuitive from the side view at the bottom of each STM diagram where the atoms with the highest position were marked by orange triangles. For the HS doping strategy, the situation is opposite: except for those in $Si_{0.875}Ge_{0.125}M$ and $Si_{0.125}Ge_{0.875}M$, the brightness of the atoms in all the others is uniform. The results indicate that the SiGeM in the HS doping case are generally smoother than those in the LE doping case. This may provide some guidance for future experimental characterizations.

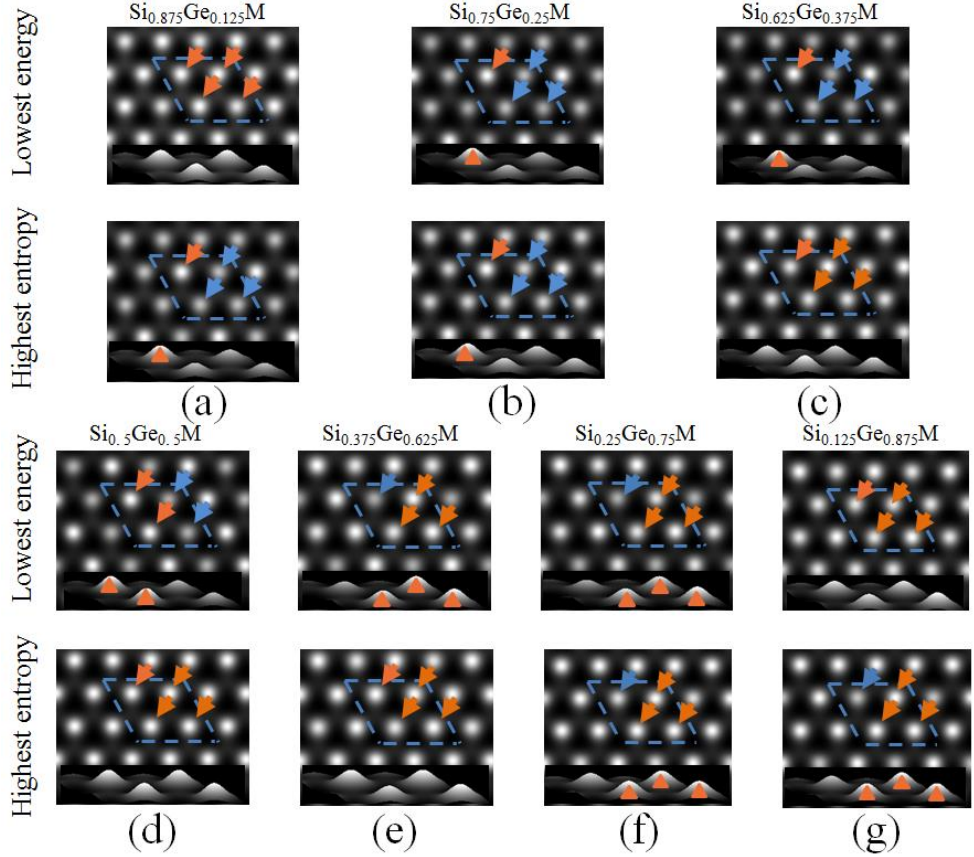

**FIG. 3.** The simulated STM images of $Si_{1-x}Ge_xM$. (a-g) The STM images for the relaxed $Si_{0.875}Ge_{0.125}M$, $Si_{0.75}Ge_{0.25}M$, $Si_{0.625}Ge_{0.375}M$, $Si_{0.5}Ge_{0.5}M$, $Si_{0.375}Ge_{0.625}M$, $Si_{0.25}Ge_{0.75}M$ and $Si_{0.125}Ge_{0.875}M$ with LE and HS doping strategies after smearing, respectively. The supercells in the STM images are highlighted with blue rhombus. Orange and blue arrows indicate relatively high and low positions in the same layer, and an orange triangle represents an atom at a higher relative position.

Raman spectroscopy has been used in the analysis of characteristic vibrational modes of 2D materials. For instance, the calculated G-like and D peaks are likely to be fingerprints of Raman spectra of silicene and germanene,[42] and the Raman frequency shift has been applied to the quantitative analysis of $MoS_2$ layers.[49] In Table 2, we show typical calculated results of Raman and infrared activities of the $Si_{1-x}Ge_xM$. The results of silicene and germanene in the table are consistent with previous work. [42] For $Si_{0.5}Ge_{0.5}M$ with *P3m1* symmetry, there are 6 phonon modes divided into 3 acoustic and 3 optical modes. Brillouin zone-center optical phonon modes shows following irreducible representation characteristics: $\Gamma_{optic}=A_1+E$, where $A_1$ mode comes from the vibrations of associated atoms along the out-of-plane direction, and $E$ mode is doubly degenerated and indicates lack of vibrations along the out-of-plane direction. Both $A_1$ and $E$ modes are Raman and infrared active. For $Si_{0.75}Ge_{0.25}M$ and $Si_{0.25}Ge_{0.75}M$ with *P-3m1* symmetry, there are 24 phonon modes divided into 3 acoustic and 21 optical modes. Brillouin zone-center optical phonon modes show the following irreducible representation characteristics: $\Gamma_{optic}=3A_{1g}+4E_g+2A_{2u}+3E_u+A_{1u}+A_{2g}$, where $A_{1g}$, $A_{2g}$, $A_{1u}$ and $A_{2u}$ modes come from the vibrations of associated atoms along the out-of-plane direction, and $E_g$ and $E_u$ modes are doubly degenerated and come from the vibrations along the in-plane direction. $A_{1g}$ and $E_g$ modes are Raman active, and $A_{2u}$ and $E_u$ modes are infrared active. It is found that the energies of the optical and acoustic phonon curves of the compounds decrease with the increase of the concentration of Ge atoms, because the strength of the interatomic bonds decreases as well.

**TABLE 2.** The typical calculated Raman and infrared active modes and Raman peak frequencies of silicene, $Si_{0.75}Ge_{0.25}M$, $Si_{0.5}Ge_{0.5}M$ and $Si_{0.25}Ge_{0.75}M$ and germanene, respectively.

| Modes | Ramman/Infrared activity | | Silicene (P-3m1) | $Si_{0.75}Ge_{0.25}M$ (P-3m1) | $Si_{0.5}Ge_{0.5}M$ (P3m1) | $Si_{0.25}Ge_{0.75}M$ (P-3m1) | Germanene (P-3m1) |
|---|---|---|---|---|---|---|---|
| | Ramman activity | Infrared activity | Calculated frequency(cm$^{-1}$) | Calculated frequency(cm$^{-1}$) | Calculated frequency(cm$^{-1}$) | Calculated frequency(cm$^{-1}$) | Calculated frequency(cm$^{-1}$) |
| $A_{1g}$ | Y | N | 183.8 | 73 | | 84.5 | 165.4 |
| $E_g$ | Y | N | 568.8 | 75 | | 64.5 | 307.1 |
| $A_{2u}$ | N | Y | | 73.1 | | 85.2 | |
| $E_u$ | N | Y | | 101.6 | | 60.1 | |
| $A_{2g}$ | N | N | | 103.6 | | 54.4 | |
| $E_g$ | Y | N | | 108.1 | | 67.9 | |
| $A_{1g}$ | Y | N | | 193.5 | | 182.6 | |
| $E_u$ | N | Y | | 331.3 | | 257.5 | |
| $E_g$ | Y | N | | 378 | | 281.6 | |
| $A_{2u}$ | N | Y | | 391.2 | | 272.3 | |
| $A_{1g}$ | Y | N | | 449 | | 281 | |
| $E_u$ | N | Y | | 482.8 | | 416.1 | |
| $A_{1u}$ | N | N | | 529.6 | | 295.4 | |
| $E_g$ | Y | N | | 532.7 | | 418.3 | |
| $A_1$ | Y | Y | | | 188.8 | | |
| $E$ | Y | Y | | | 443.7 | | |

Since the SiGeM deviates from planar geometry and the buckling degree of the SiGeM increases compared to that of graphene, the effective spin-orbit coupling (SOC) will also increase,[6] implying that quantum spin hall effect (QSHE) of the SiGeM will be more significant. Additionally, since a Ge atom has larger intrinsic SOC strength than a Si one does, the naïve speculation would be that SOC will increase as the Ge concentration increases. Therefore, the band structures of the SiGeM with different Ge concentrations were calculated by PBE + SOC corrections.

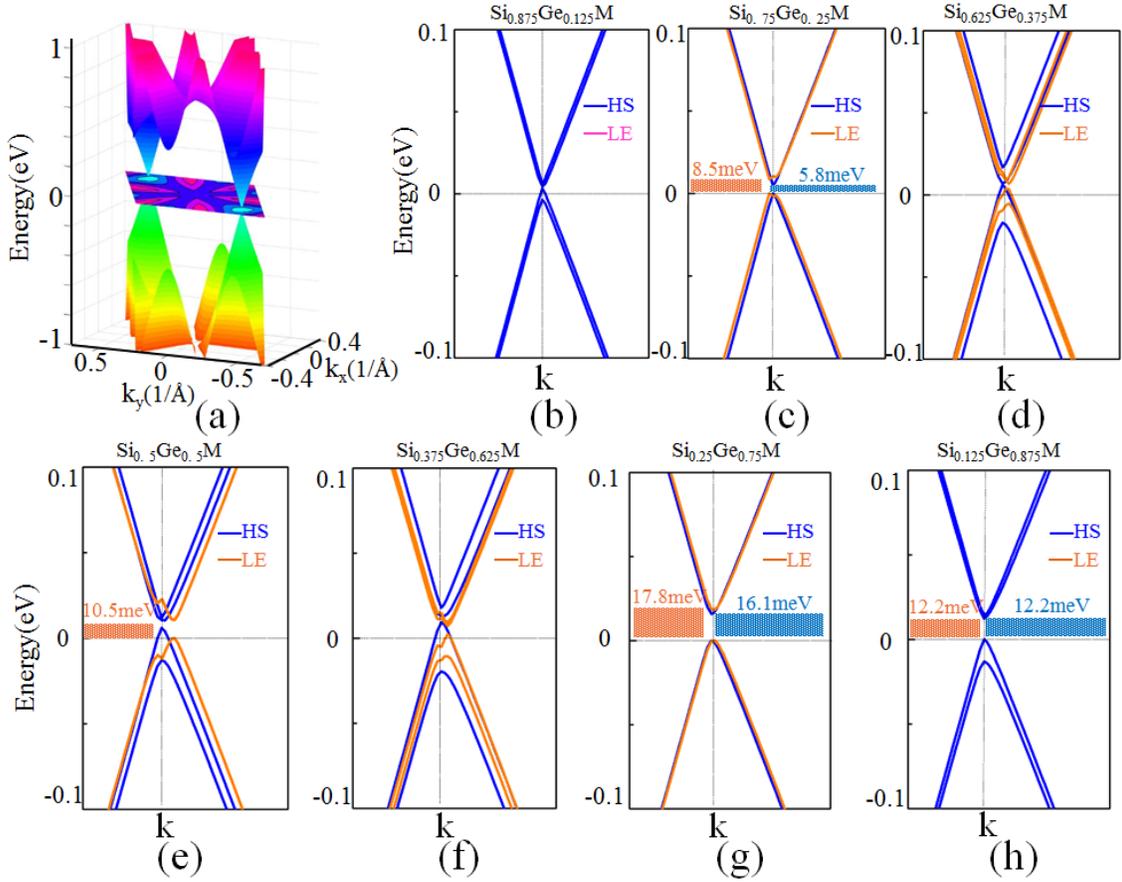

**FIG. 4.** The band structures and band gap values of SiGeM. (a) The 3D band structure of $Si_{1-x}Ge_xM$ (x = 0.25) in color form. (b-h) The band structure diagrams of $Si_{0.875}Ge_{0.125}M$, $Si_{0.75}Ge_{0.25}M$, $Si_{0.625}Ge_{0.375}M$, $Si_{0.5}Ge_{0.5}M$, $Si_{0.375}Ge_{0.625}M$, $Si_{0.25}Ge_{0.75}M$ and $Si_{0.125}Ge_{0.875}M$ with HS and LE doping strategies, respectively. All of them are marked with red circle in Figure 1. The Fermi level or valence band top is set as zero.

Figure 4a gives a typical 3d band structure of $Si_{1-x}Ge_xM$ (x = 0.25) without considering SOC, which shows two distinct Dirac cones formed at K and K' points. Due to the symmetry at K and K', the ambipolar character in the small energy range near the $E_F$ is very significant. Around the crossing point, these bands are linear, therefore charge carriers behave like massless Dirac fermions, which could be useful for electronic applications. Figure 4b-h give the band structures

of $Si_{0.875}Ge_{0.125}M$, $Si_{0.75}Ge_{0.25}M$, $Si_{0.625}Ge_{0.375}M$, $Si_{0.5}Ge_{0.5}M$, $Si_{0.375}Ge_{0.625}M$, $Si_{0.25}Ge_{0.75}M$ and $Si_{0.125}Ge_{0.875}M$ with SOC in the LE and HS doping cases. For the convenience of comparison, we also calculated the band structures of silicene and germanene under SOC. The band gap widths of silicene (4meV) and germanene(23.6meV) are basically the same as those in previous work.[6,50] Interestingly, when the SOC is taken into account, the existence of low buckling structure can open an obvious gap at Dirac point. In the LE doping case, $Si_{0.75}Ge_{0.25}M$, $Si_{0.5}Ge_{0.5}M$, $Si_{0.25}Ge_{0.75}M$, and $Si_{0.125}Ge_{0.875}M$ show direct band gaps of 8.5meV, 10.5meV, 17.8meV and 12.2meV, respectively. In the HS doping case, $Si_{0.75}Ge_{0.25}M$, $Si_{0.25}Ge_{0.75}M$ and $Si_{0.125}Ge_{0.875}M$ show direct band gaps of 5.8meV, 16.1meV and 12.2meV, respectively. Therefore, SOC in the low-buckled structures can achieve non-trivial band topological properties in the SiGeM, indicating that QSHE could be realized in the materials.

The diagrams of band gap values vs. Ge concentrations in the $Si_{1-x}Ge_xM$ with the LE and HS doping strategies were shown in Figure 4. In both strategies, the band gap values change in oscillatory ways as Ge concentration x changes. Interestingly, when x equals to 0.5, the LE doping strategy results in a narrow band gap semiconductor, while the HS doping strategy results in a metal with zero band gap. The results probably result from the interplay between SOC and crystalline field splitting: as x changes, both the buckling degree ($\triangle$) and space symmetry of the monolayer changes, and $\triangle$-dependent SOC and symmetry-dependent crystalline field splitting change, but in different ways, resulting in oscillatory variations of band gap values. In addition, since different symmetries in the LE and HS doped samples generate different crystalline fields and influence band structures differently, different variation of band structures are obtained in different doping strategies. Especially, two different electronic states (semiconductor and metal) are obtained at 50% doping concentration for LE and HS doping strategies, respectively.

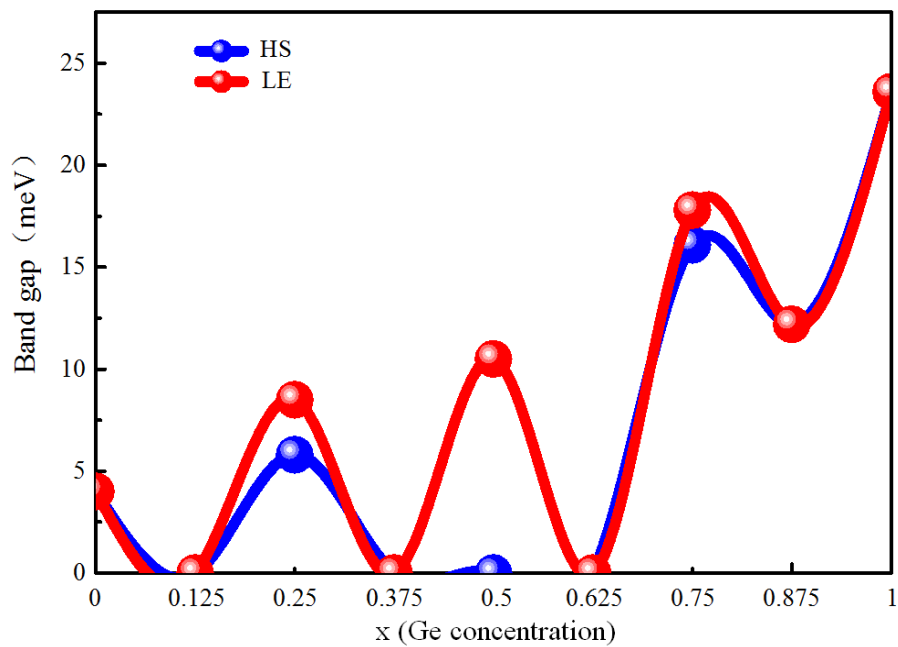

**FIG. 5.** The band gap values of the Si$_{1-x}$Ge$_x$M as the function of Ge concentration x with the LE and HS doping strategies, respectively.

Based on the diagram in Figure 5, an alternative method of tuning the band gap values of the 2D materials via varying doping concentrations and doping strategies can be proposed. This method is different from previously reported methods used in other 2D systems such as stress modification, element adsorption and electric/magnetic field tuning,[51-55] and is robust since the elementary concentration x can vary from 0 to 100% and the band gaps can be well controlled in a non-trivial way. As a result, the SiGeM can change between Dirac metals and narrow-band semiconductors, depending on the doping concentrations and doping strategies. We note that it has been shown very recently that the band gaps of germanene can be tuned to ~1.5 eV by hydrogen functionalization.[25] Combing engineering of doping concentrations and strategies and other techniques such as hydrogen functionalization, SiGeM could show great electronic and optical tunability for design and application of 2D semiconducting materials.

**TABLE 3.** The average effective masses at VBM ($m_h$) and CBM ($m_e$) in the semiconducting SiGeM in the LE and HS doping cases, respectively. $m_0$ denotes the mass of free electrons.

| | Name | Silicene | $Si_{0.75}Ge_{0.25}M$ | $Si_{0.5}Ge_{0.5}M$ | $Si_{0.25}Ge_{0.75}M$ | $Si_{0.125}Ge_{0.875}M$ | Germanene |
|---|---|---|---|---|---|---|---|
| LE | $m_h(m_0)$ | 0.0208 | 0.013 | 0.0146 | 0.0088 | 0.0083 | 0.0085 |
| | $m_e(m_0)$ | 0.021 | 0.0132 | 0.0144 | 0.0086 | 0.01 | 0.0083 |
| HS | $m_h(m_0)$ | 0.0208 | 0.0128 | | 0.0138 | 0.0083 | 0.0085 |
| | $m_e(m_0)$ | 0.021 | 0.0128 | | 0.0129 | 0.01 | 0.0083 |

Finally, the effective masses of the carriers in the semiconducting SiGeM were studied. Here, the equation of $1/m = (\partial^2 E(k))/(\hbar^2 \partial k^2)$ is used to obtain the effective masses of the valence band maximum (VBM) and the conduction band minimum (CBM) with considering the influence of SOC, where $\hbar$ is the reduced Planck constant. As shown in Table 3, the effective masses of the carriers are between $0.08m_0$ and $0.21m_0$ in either the LE or HS case, comparable to those of bilayer graphene, silicene and germanene.[56,57] We note that the effective masses in silicene ($m_h$ = 0.021 $m_0$, $m_e$ = 0.021 $m_0$) and germanene ($m_h$ = 0.009 $m_0$, $m_e$ = 0.008 $m_0$) obtained in our work are consistent with previous theoretical work[57] within the accuracy limits of DFT. The low effective masses of the carriers suggest the potentials of the SiGeM in the applications of fast and low energy cost electronic devices.

## CONCLUSIONS

In summary, we have systematically investigated the stabilities and structural and electronic properties of $Si_{1-x}Ge_xM$ (0<x<1). Our results reveal a structurally stable 2D binary system in which the two elements are infinitely miscible with each other. The structural characteristics was explored further by simulated STM images and Raman and infrared peaks. The study of electronic properties reveals oscillatory behaviors of band gap variations as x changes and low effective masses of carriers in the $Si_{1-x}Ge_xM$. The $Si_{1-x}Ge_xM$ with stable structures and excellent properties would be provide varieties in structure and functionalities and be important for 2D applications of group IV semiconductors.

ACKNOWLEDGMENT

This work was supported by National Key R&D Plan of China through Grant No. 2017YFB0405702 and National Natural Science Foundation of China through Grant No. 51672179.